\documentclass[%
 reprint,
 superscriptaddress,
 amsmath,amssymb,
 aps,prx
]{revtex4-1}

\usepackage[utf8]{inputenc}
\usepackage[T1]{fontenc}
\usepackage{graphicx}
\usepackage{dcolumn}
\usepackage{bm}
\usepackage{nicefrac, xfrac}
\usepackage[dvipsnames]{xcolor}
\usepackage{subcaption}
\usepackage{float}
\usepackage[caption=false]{subfig}
\usepackage{soul}

\begin{document}

% \preprint{APS/123-QED}
\title{Symmetry-guided Design Principles for Spin Splitting and Hall Transport in Orthorhombic Altermagnetic Perovskites
} 
\author{Rasmita Kumari}
\affiliation{Department of Physics, Indian Institute of Technology Bombay, Powai, Mumbai 400076, India}
\author{Bishal Das}
\affiliation{Department of Physics, Indian Institute of Technology Bombay, Powai, Mumbai 400076, India}

\author{Aftab Alam}
\email{aftab@iitb.ac.in}

\affiliation{Department of Physics, Indian Institute of Technology Bombay, Powai, Mumbai 400076, India}

\begin{abstract}

Magnetic symmetry can generate momentum-dependent spin-split electronic bands even in the absence of spin--orbit coupling (SOC), giving rise to the recently discovered class of altermagnets. Here, we establish a unified symmetry framework for collinear antiferromagnets in orthorhombic $Pbnm$ perovskites by combining spin and magnetic group theory with first-principles calculations. We show that the irreducible representation of the magnetic order uniquely determines the momentum-space planes supporting altermagnetic spin splitting, the form of the effective low-energy Hamiltonian, the orientation of SOC-induced weak ferromagnetic canting, and the allowed anomalous Hall conductivity (AHC) tensor components. These predictions are validated in eight experimentally realized orthorhombic perovskite oxides spanning both insulating and metallic regimes. In particular, LaTiO$_3$ and CaCrO$_3$ exhibit sizable anomalous Hall conductivities of approximately 38 and 205~S/cm, respectively, despite nearly vanishing net magnetization. We further show that SOC gaps symmetry-protected altermagnetic band crossings, generating large Berry curvature, while the Hall response is fundamentally rooted in the underlying non-relativistic spin splitting. Our work establishes a predictive symmetry-based framework for discovering and engineering altermagnetic materials with tunable spin-dependent electronic and transport properties.
% xf

\end{abstract}
\maketitle
\section{Introduction}

The study of spin splitting in electronic band structures has emerged as a central topic in condensed matter physics because of its fundamental significance and its potential for next-generation spintronic applications. Traditionally, magnetic materials are broadly classified into ferromagnets (FMs) and antiferromagnets (AFMs). In ferromagnets, the breaking of time-reversal symmetry produces net spin polarization, enabling efficient spin manipulation. However, the accompanying stray magnetic fields interfere with neighboring devices and limit scalability. In addition, the spin dynamics in ferromagnets are typically restricted to the GHz regime. In contrast, antiferromagnets possess fully compensated magnetic moments and can exhibit ultrafast spin dynamics reaching the THz regime owing to their strong exchange interactions~\cite{Jungwirth2016, Baltz2018}. These advantages have motivated extensive efforts to explore AFM materials for spintronic technologies.

Conventionally, spin splitting in band structures has been associated with relativistic spin-orbit coupling (SOC) effects such as the Rashba and Dresselhaus mechanisms~\cite{Bychkov1984, Dresselhaus1955}. These effects require broken inversion symmetry and are generally prominent in compounds containing heavy elements~\cite{Manchon2015}. However, the reliance on strong SOC not only limits the range of suitable materials but also weakens chemical bonding in many systems. More importantly, in conventional collinear AFMs, the combined symmetries of time reversal and lattice operations enforce spin degeneracy throughout the Brillouin zone, thereby suppressing spin splitting and limiting their direct applicability in spintronics. These limitations have stimulated the search for unconventional magnetic states that simultaneously host compensated magnetism and spin-split electronic bands.

A major breakthrough in this direction was the recent discovery of nonrelativistic spin splitting in collinear antiferromagnets, leading to the emergence of a new magnetic phase termed altermagnetism~\cite{Smejkal2022Nature}. The concept was first theoretically proposed in RuO$_2$~\cite{Smejkal2020RuO2}, where signatures such as nonrelativistic spin splitting and anomalous Hall conductivity were predicted. Although the precise magnetic ground state of RuO$_2$ remains under debate~\cite{Maznichenko2021}, subsequent theoretical and experimental studies on materials including $\alpha$-MnTe, CrSb, GdAlSi, Mn$_5$Si$_3$, and MnTe$_2$~\cite{Naka2019, Samanta2020, Gonzalez2021, Nag2024GdAlSi} have firmly established altermagnetism as a distinct class of compensated magnetic order. More recently, experimental evidence for two-dimensional altermagnetism has also been reported in Rb$_{1-\delta}$V$_2$Te$_2$O~\cite{Cui2023}.

The defining characteristic of altermagnetism is that opposite-spin sublattices are not connected by either inversion or primitive translational symmetries~\cite{Smejkal2022PRX}. They are rather connected through rotational, mirror, glide, or screw symmetries~\cite{Hayami2021}, giving rise to momentum-dependent spin splitting even in the absence of SOC.
Motivated by these unusual symmetry properties, several theoretical studies have investigated the electronic structure and transport responses of altermagnets. For example, the band structures near the $\Gamma$ and $A$ points in $\alpha$-MnTe have been successfully described using effective Hamiltonian models~\cite{Takahashi2025MnTe}. Unconventional spin Hall conductivity arising from the interplay between magnetic and crystalline symmetries has been reported in RuO$_2$, $\alpha$-MnTe and CrSb~\cite{Jeong2025USHC}. Symmetry breaking has also been shown to induce characteristic spin-orbital textures and weak spin magnetization~\cite{Wang2024SpinOrbitalAltermagnetism}. In addition, a general framework for stabilizing two-dimensional altermagnetism through symmetry-protected coincident van Hove singularities has recently been proposed~\cite{Yu2025}. Nonrelativistic spin splitting has further been demonstrated in collinear antiferromagnetic 6H perovskites, where layered structural motifs break $\bar{E}\tau$ symmetry~\cite{Streltsov2025Perovskite}, where $\bar{E}$ and $\tau$ stands for inversion and time reversal symmetry respectively. Moreover, anomalous Hall conductivity in altermagnets has been shown to be governed primarily by symmetry and Néel vector orientation rather than by any net ferromagnetic moment~\cite{Smejkal2020CrystalTR, Roig2024Quasisymmetry}.

Among the broad classes of crystalline materials, perovskites provide an exceptionally versatile platform for realizing unconventional magnetic phases because they encompass metals, semiconductors, and insulators with highly tunable structural, electronic, and magnetic degrees of freedom~\cite{maekawa2004physics, Imada1998}. In altermagnets, time-reversal symmetry breaking is influenced not only by magnetic ordering but also by the surrounding nonmagnetic sublattice environment. This strong coupling between lattice symmetry and magnetism makes perovskites particularly promising candidates for hosting altermagnetic states. Orthorhombic perovskites with magnetic propagation vector $\mathbf{k}=0$ are especially intriguing because their primitive magnetic unit cell coincides with the primitive nonmagnetic unit cell, naturally eliminating the translational symmetry that usually connects opposite-spin sublattices. Consequently, the spin sublattices are related primarily through rotational symmetries, providing a favorable setting for d-wave altermagnetism.

Motivated by these considerations, in this paper, we investigate a family of orthorhombic perovskites exhibiting different antiferromagnetic orders and analyze their altermagnetic properties using symmetry-based approaches. Through detailed symmetry analysis, we identify the momentum-space planes where spin splitting is allowed in both nonrelativistic and relativistic regimes and construct the corresponding effective Hamiltonians around the $\Gamma$ point. To explore the potential relevance of these systems for spintronics, we further calculate the anomalous Hall conductivity in metallic CaCrO$_3$ and the Mott-insulating compound LaTiO$_3$.
Interestingly, we note that previous studies on altermagnetism in the $Pnma$ structure contain certain ambiguities regarding the correspondence between crystallographic and global coordinate axes~\cite{Rooj2025}. Furthermore, the role of spin orientations in magnetic space groups in the presence of SOC was not explicitly addressed. In the present work, we carefully account for these aspects to provide a consistent and unambiguous symmetry description of orthorhombic altermagnetic perovskites.

\section{Computational method Details}

Density functional theory (DFT) calculations~\cite{Hohenberg1964,Kohn1965} were performed using the Vienna \emph{ab-initio} Simulation Package (VASP)~\cite{Kresse1996a,Kresse1996b} within the projector augmented-wave (PAW) formalism~\cite{Blochl1994}. Exchange--correlation effects were treated using the Perdew--Burke--Ernzerhof (PBE) functional within the generalized gradient approximation (GGA), together with the Hubbard $U$ correction~\cite{Dudarev1998}. The effective on-site Hubbard parameters were chosen as follows: $U = 6$ eV for the La-$5d$ orbitals, $U = 2.5$ eV for the Mn-$3d$, Fe-$3d$, and Cr-$3d$ orbitals in LaMO$_3$ ($M =$ Mn, Fe, Cr)~\cite{Okugawa2018}, and $U = 2.3$ eV for the Ti-$3d$ orbitals in LaTiO$_3$~\cite{Maznichenko2024}. Also, $U = 2.5$ eV for the Cr-$3d$ orbitals in YCrO$_3$, $U = 1.9$ eV for the Ni-$3d$ orbitals in SeNiO$_3$~\cite{Munoz2006}, $U = 5$ eV for the Mn-$3d$ orbitals in SeMnO$_3$~\cite{Munoz2006}, and $U = 3.0$ eV for the Cr-$3d$ orbitals in CaCrO$_3$. These $U$ values are calculated self-consistently for each compounds.  All the calculations were performed using experimentally reported crystal structures~\cite{ICSD}.  The corresponding calculated properties of these compounds are summarized in Table~\ref{tab:allmaterials}. A $\Gamma$-centered $6 \times 6 \times 5$ Monkhorst--Pack $k$-point mesh~\cite{Monkhorst1976}, a plane-wave energy cutoff of 520 eV, and an electronic self-consistency convergence criterion of $10^{-5}$ eV were employed throughout.

For LaTiO$_3$ and CaCrO$_3$, maximally localized Wannier functions were constructed using the \textsc{Wannier90} package~\cite{Mostofi2008}. The Bloch states were projected onto localized atomic orbitals consisting of La ($s$, $p$, $d$), Ti ($s$, $p$, $d$), and O ($s$, $p$) orbitals for LaTiO$_3$, and Ca ($s$, $p$), Cr ($s$, $p$, $d$), and O ($s$, $p$) orbitals for CaCrO$_3$. The anomalous Hall conductivity (AHC)~\cite{Fukui2005} was subsequently calculated using the \textsc{WannierTools} package on a dense $120 \times 120 \times 100$ $k$-point mesh.

\section{\protect\label{sec:results-and-discussion}results and discussion}

\subsection{$Pbnm$ Perovskite Oxides (ABO$_3$)}

Perovskite oxides of the ABO$_3$ type provide an ideal platform for realizing a wide variety of magnetic ground states because of their structural flexibility and rich interplay between lattice, orbital, and spin degrees of freedom. Their crystal structure consists of corner-sharing BO$_6$ octahedra, while the larger A-site cations occupy the interstitial positions between the octahedra~\cite{ref1}. The structural stability of these compounds is commonly characterized by the Goldschmidt tolerance factor, which relates the ionic radii of the constituent atoms. When the tolerance factor is close to unity, the ideal cubic structure is stabilized, whereas deviations from unity generally induce lattice distortions~\cite{ref2}. The most prominent distortions include GdFeO$_3$-type octahedral tilting~\cite{ref3,ref4} and Jahn--Teller distortions, which remove orbital degeneracies through alternating B--O bond lengths~\cite{ref5}. In many perovskites, the cubic $Pm\bar{3}m$ phase first transforms into a tetragonal structure driven by Jahn--Teller distortion and subsequently evolves into the orthorhombic $Pbnm$/$Pnma$ phase through cooperative octahedral tilts and A-site displacements~\cite{ref6}. For larger ionic-size mismatch, these structural instabilities may cooperate to produce a direct cubic-to-orthorhombic transition~\cite{ref7}. Importantly, the resulting symmetry lowering introduces nonsymmorphic symmetries that combine point-group operations with fractional lattice translations, such as glide planes and screw axes~\cite{ref8}.

In orthorhombic perovskites, the choice of crystallographic convention is particularly important because different space-group settings are frequently employed in the literature. The standard description corresponds to the $Pbnm$ ($D_{2h}$) setting, in which the global axes $(x,y,z)$ coincide with the crystallographic axes $(a,b,c)$~\cite{Bertaut1968}. An alternative but symmetry-equivalent description is provided by the $Pnma$ setting, where the correspondence between the global and crystallographic axes is rotated such that $(x,y,z)$ maps onto $(c,a,b)$~\cite{Litvin_MagneticGroups}. Although these conventions are mathematically equivalent, the distinction becomes crucial for the correct interpretation of symmetry operations and for maintaining consistency between theoretical calculations and experimental observations. 

The orthorhombic $Pbnm$ ($D_{2h}$) structure contains four formula units within the primitive unit cell. The full space group consists of eight symmetry operations that can be generated from three independent operations: inversion symmetry $\bar{E}$, a twofold screw rotation about the $x$ axis, $S_{2x} = \{C_{2x} \,|\, (\tfrac{1}{2},0,\tfrac{1}{2})\}$, and a twofold screw rotation about the $y$ axis,
 $S_{2y} = \{C_{2y} \,|\, (\tfrac{1}{2},\tfrac{1}{2},\tfrac{1}{2})\}$~\cite{Bertaut1968}. Equivalently, the symmetry can also be described in terms of a mirror plane perpendicular to the $z$ direction, a $b$-glide plane perpendicular to $x$, and an $n$-glide plane perpendicular to $y$, which together define the nonsymmorphic character of the $Pbnm$ structure~\cite{Bertaut1968}.

\begin{figure*}[t]
    \centering
    \includegraphics[
        width=1.0\linewidth]{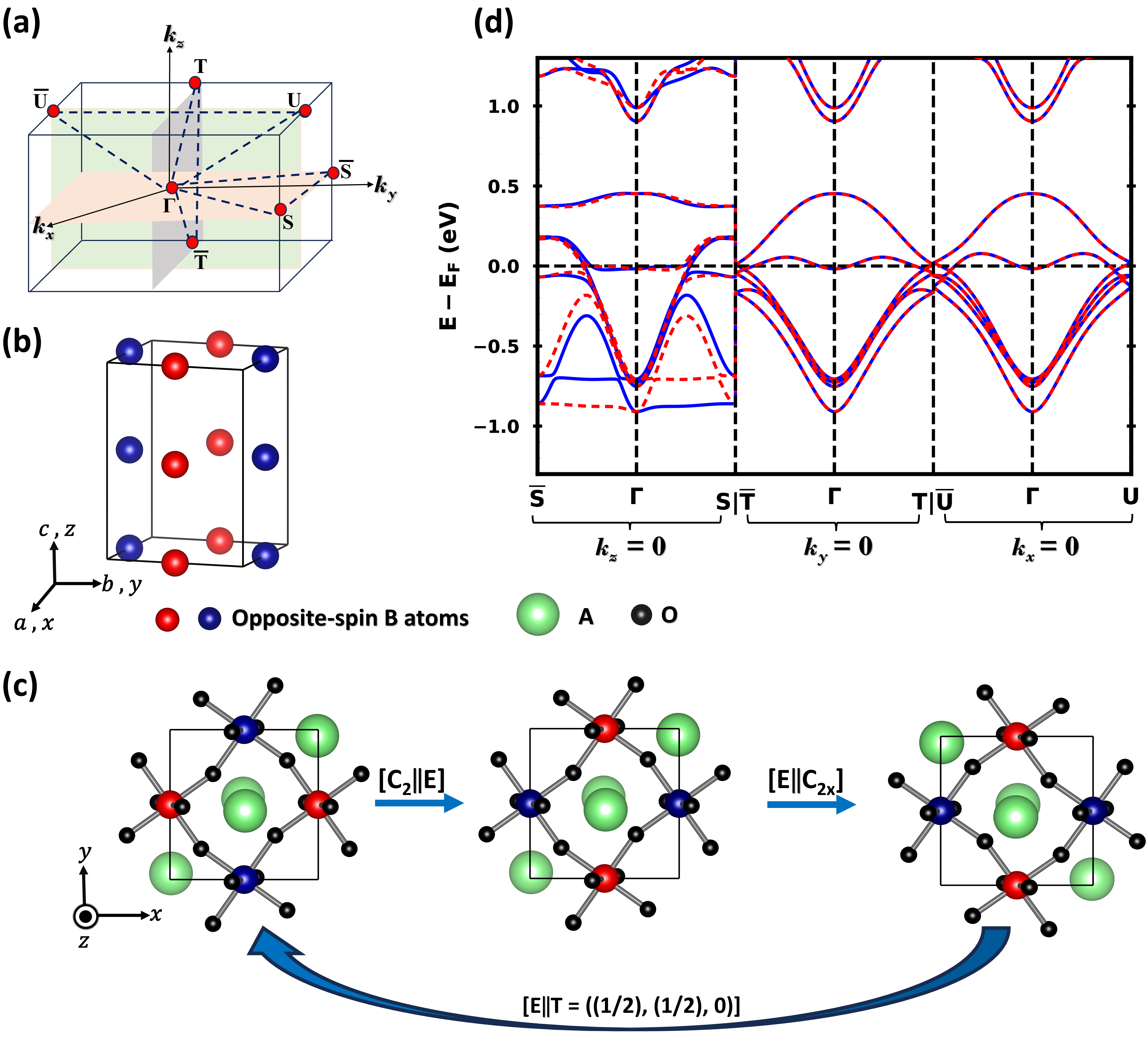}
    \caption{(a) High-symmetry points and paths in the orthorhombic Brillouin zone of ABO$_3$ used for the symmetry analysis. (b) C-type antiferromagnetic (AFM) spin configuration with opposite-spin sublattices shown in red and blue. (c) Symmetry relation between opposite-spin sublattices, connected by a spin-space $C_2$ rotation together with a real-space $C_{2x}$ screw operation and fractional translation $\mathbf{T}=(\frac12,\frac12,0)$. (d) Spin-polarized DFT band structure of C-type AFM CaCrO$_3$ without SOC. Spin splitting occurs exclusively in the symmetry-allowed $k_z=0$ plane, whereas the $k_x=0$ and $k_y=0$ planes remain spin degenerate. Red and blue denote bands belonging to opposite-spin sublattices. } 
    \label{fig:CCC-AFMMM}
\end{figure*}

% \bd{The following paragraph was not printed before due to syntax errors.}
When antiferromagnetic order is introduced, the magnetic primitive cell coincides with the nonmagnetic primitive cell. As a result, the primitive translation that would otherwise connect opposite-spin sublattices is no longer a symmetry operation. Since inversion symmetry leaves the spin vectors invariant, the resulting magnetic phase belongs to the third magnetic class, namely altermagnetism.
To understand the origin of spin splitting in antiferromagnetic phase of these systems, we construct an effective Hamiltonian constrained by the underlying magnetic symmetries, as discussed in following subsections for each type of AFM ordering. At the $\Gamma$ point, the Hamiltonian remains invariant under the full set of symmetry operations, making it a natural reference point for symmetry analysis. Away from $\Gamma$, however, the little groups contain fewer symmetry operations, and the absence of symmetry relations connecting opposite-spin sublattices permits additional symmetry-allowed terms that generate spin splitting. To systematically analyze these effects, we examine the residual symmetries associated with the parent space group in three representative high-symmetry planes: $k_z = 0$, $k_y = 0$, and $k_x = 0$. The corresponding high-symmetry paths are $\bar{S} \rightarrow \Gamma \rightarrow S$ [(-0.5, 0.5, 0) $\rightarrow$ (0, 0, 0) $\rightarrow$ (0.5, 0.5, 0)], $\bar{T} \rightarrow \Gamma \rightarrow T$ [(-0.5, 0, -0.5) $\rightarrow$ (0, 0, 0) $\rightarrow$ (-0.5, 0, 0.5)], and $\bar{U} \rightarrow \Gamma \rightarrow U$ [(0, -0.5, 0.5) $\rightarrow$ (0, 0, 0) $\rightarrow$ (0, 0.5, 0.5)]  respectively, as shown in Fig.~\ref{fig:CCC-AFMMM}(a). Using these symmetry constraints, we derive the corresponding effective Hamiltonians and validate the resulting predictions through \textit{ab-initio} band-structure calculations for three types of antiferromagnetic orders i.e.  A-, C-, and G-type, as discussed below.

\subsection{\textbf{Antiferromagnetic symmetry without spin–orbit coupling}}
\subsubsection{\textbf{C-Type AFM Symmetry}}

The C-type antiferromagnetic (C-AFM) order consists of ferromagnetically aligned spins along the $c$ axis, while neighboring spins within the $ab$ plane are coupled antiferromagnetically (see Fig.~\ref{fig:CCC-AFMMM}(b)). This layered magnetic arrangement lowers the symmetry of the parent $Pbnm$ structure and gives rise to the magnetic space group $Pn'ma'$ or $Pb'n'm$~\cite{Burns1977, Okugawa2018}.
Let us first analyze the non-relativistic electronic structure using the spin point-group formalism. In the absence of spin--orbit coupling (SOC), the Hamiltonian is governed solely by the anisotropy field and Heisenberg exchange interactions~\cite{BrinkmanElliott1966}. Under these conditions, the relevant symmetry operations belong to spin groups, in which real-space and spin-space transformations are decoupled and represented in the form $[R_i \parallel R_j]$, where the left (right) operation acts in spin (real) space. Accordingly, the nonrelativistic spin splitting in collinear compensated magnetic systems is naturally described using spin-group symmetries, whereas the relativistic case with SOC is more appropriately treated within the magnetic-group formalism.

From the spin point-group perspective, the C-AFM phase belongs to the group ${}^{2}m{}^{2}m{}^{1}m$~\cite{Smejkal2022PRXX}. The residual spatial point-group symmetries form a $2/m$ halving subgroup containing four elements: $\{E, \bar{E}, C_{2z}, m_z\}$, where $E$ denotes the identity operation. These operations connect sites with the same spin orientation, while the remaining coset elements connect opposite-spin sublattices. As illustrated in Fig.~\ref{fig:CCC-AFMMM}(c), the opposite-spin sublattices are related by a $C_{2x}$ rotation combined with a fractional translation $\mathbf{T} = (\tfrac{1}{2},\tfrac{1}{2},0)$ along the $(110)$ direction. Consequently, the operation $C_{2x}$ exchanges the two spin sublattices, and the remaining coset elements can be generated through the product $[C_2 \parallel C_{2x}]$. The resulting nontrivial spin Laue group $R^{\mathrm{III}}_s$ is therefore expressed as
\[
{}^{2}m{}^{2}m{}^{1}m
=
[E \parallel 2/m]
+
[C_2 \parallel C_{2x}]
[E \parallel 2/m],
\]
which is characterized by a spin-group integer equal to $2$.

\begin{table*}[t]
\centering
\caption{Transformation properties of the spin state under the spin point group in the absence of SOC.}
\label{tab:sz_transform}

\small
\renewcommand{\arraystretch}{1.25}
\setlength{\tabcolsep}{7pt}   % Increase horizontal padding

\begin{tabular}{|c|c|c|c|c|c|c|c|c|}
\hline
\shortstack{\textbf{Spin transformation}} &
\textbf{$[E\Vert E]$} &
\textbf{$[E\Vert \bar{E}]$} &
\textbf{$[E\Vert C_{2x}]$} &
\textbf{$[E\Vert C_{2y}]$} &
\textbf{$[E\Vert C_{2z}]$} &
\textbf{$[C_{2}\Vert C_{2x}]$} &
\textbf{$[C_{2}\Vert C_{2y}]$} &
\textbf{$[C_{2}\Vert C_{2z}]$} \\
\hline

$s_z$ &
1 &
1 &
1 &
1 &
1 &
$-1$ &
$-1$ &
$-1$ \\
\hline

\end{tabular}
\end{table*}

\begin{table}
\centering
\caption{Transformation properties of irreducible momentum tensors up to second order at the $\Gamma$-point under the generators of the spin point group in the absence of SOC for the orthorhombic C-AFM phase.}
\label{tab:c-afm-spin}

\small
\renewcommand{\arraystretch}{1.25}

\resizebox{\columnwidth}{!}{%
\begin{tabular}{|c|c|c|c|c|}
\hline
\textbf{Irreducible Tensor} &
\textbf{$[E\Vert E]$} &
\textbf{$[E\Vert \bar{E}]$} &
\textbf{$[E\Vert C_{2z}]$} &
\textbf{$[C_{2}\Vert C_{2y}]$} \\
\hline

$k_x^2,\;k_y^2,\;k_z^2$
& 1 & 1 & 1 & 1 \\
\hline

$k_xk_y$
& 1 & 1 & 1 & $-1$ \\
\hline

$k_yk_z$
& 1 & 1 & $-1$ & $-1$ \\
\hline

$k_zk_x$
& 1 & 1 & $-1$ & 1 \\
\hline

\end{tabular}
}
\end{table}

We next examine how these symmetry operations constrain spin degeneracy in momentum space. Since the present symmetry analysis concerns the transformation of the momentum basis functions at the Brillouin-zone center ($\Gamma$-point), the translational part of the nonsymmorphic symmetry operations contributes only the trivial phase factor, $e^{i\mathbf{k}\cdot\mathbf{T}} = 1$. Therefore, the transformation properties are completely determined by the corresponding point-group operations, which are used throughout the following symmetry tables.

As the halving subgroup corresponds to the ferromagnetically aligned sublattice, little groups containing only these operations cannot enforce spin degeneracy. Therefore, spin degeneracy can survive only when the little group contains at least one coset element connecting opposite-spin sublattices. Along the $k_x=0$ plane, the little group contains the operation $C_{2x}$, which exchanges opposite-spin sublattices. Similarly, along the $k_y=0$ plane, the corresponding role is played by $C_{2y}$. In contrast, the little group associated with the $k_z=0$ plane contains no symmetry operation that connects opposite-spin sublattices. Consequently, spin degeneracy is not protected on this plane, allowing spin splitting to emerge. This prediction is fully consistent with the calculated band structure shown in Fig.~\ref{fig:CCC-AFMMM}(d).

To construct the effective Hamiltonian near the $\Gamma$ point, we retain only those terms that remain invariant under all symmetry operations of the C-AFM phase. 

According to representation theory~\cite{Webb2003}, a term is symmetry allowed only if it transforms according to a representation containing the identity representation. 

Table~\ref{tab:sz_transform} summarizes the transformation properties of the spin state $s_z$ in the absence of SOC. The transformation properties of the irreducible momentum tensors up to second order in $\mathbf{k}$ under the generators of the spin point group are summarized in Table~\ref{tab:c-afm-spin}. Using these transformation properties together with the representation theory condition, we find that the only quadratic spin-splitting invariant permitted by symmetry is $\ s_z k_x k_y$,
which is fully consistent with the symmetry analysis discussed above.

Altogether, the symmetry classification, effective Hamiltonian analysis, and \textit{ab initio} band-structure calculations provide a consistent and unified description of spin splitting in the C-AFM phase.

\begin{figure*}[t]
    \centering
    \includegraphics[
        width=1.0\linewidth]{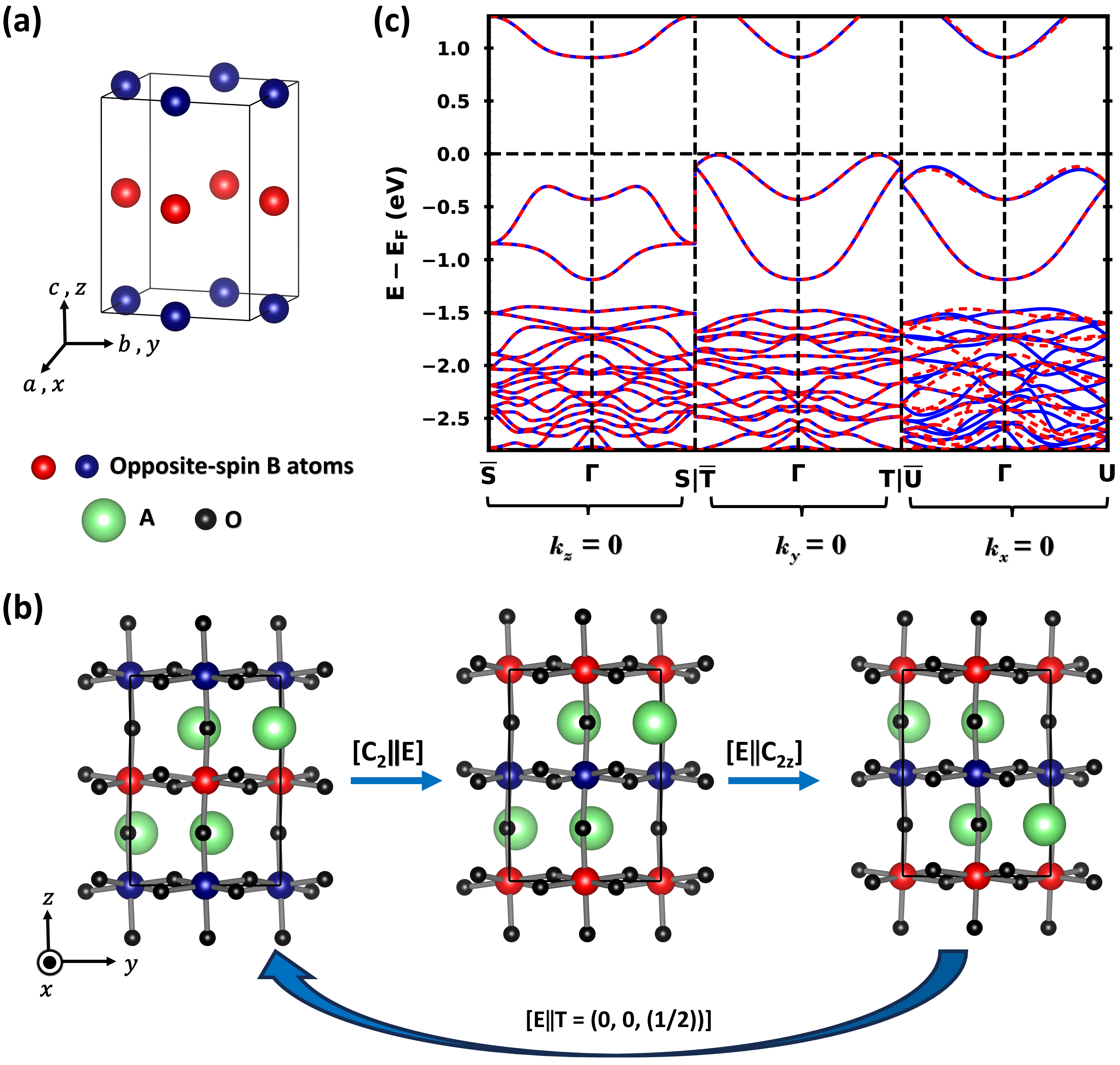}
    \caption{ (a) A-type AFM spin configuration. (b) Symmetry operation relating opposite-spin sublattices through a spin-space $C_2$ rotation followed by a real-space $C_{2z}$ screw operation and fractional translation $\mathbf{T}=(0,0,\frac12)$. (c) Spin-polarized DFT band structure of A-type AFM LaMnO$_3$ without SOC, showing symmetry-allowed spin splitting only in the $k_x=0$ plane. }
    \label{fig:AAAA_type}
\end{figure*}

\subsubsection{\textbf{A-Type AFM Symmetry}}
\begin{table}[b]
\centering
\caption{Same as Table \ref{tab:c-afm-spin}, but for the A-type AFM phase.}
\label{tab:A-TYPE}

\small
\renewcommand{\arraystretch}{1.25}

\resizebox{\columnwidth}{!}{%
\begin{tabular}{|c|c|c|c|c|}
\hline
\textbf{Irreducible Tensor} &
\textbf{$[E\Vert E]$} &
\textbf{$[E\Vert \bar{E}]$} &
\textbf{$[E\Vert C_{2x}]$} &
\textbf{$[C_{2}\Vert C_{2y}]$} \\
\hline

$k_x^2,\;k_y^2,\;k_z^2$
& 1 & 1 & 1 & 1 \\
\hline

$k_xk_y$
& 1 & 1 & $-1$ & $-1$ \\
\hline

$k_yk_z$
& 1 & 1 & 1 & $-1$ \\
\hline

$k_zk_x$
& 1 & 1 & $-1$ & 1 \\
\hline

\end{tabular}
}
\end{table}
In the A-type AFM (A-AFM) order, spins align ferromagnetically within the $ab$ plane and antiferromagnetically along the $c$ axis (Fig.~\ref{fig:AAAA_type}(a)). This magnetic structure has the same spin point-group classification as the C-type AFM state, ${}^{2}m{}^{2}m{}^{1}m$~\cite{Smejkal2022PRXX}. The residual symmetries form a $2/m$ halving subgroup, 
including $C_{2x}$ and three additional operations 
$\{E, \bar{E}, m_{x}\}$ that connect sites with the same spin orientation. The coset generator can be identified as $C_{2z}$, as illustrated in Fig.~\ref{fig:AAAA_type}(b), opposite-spin sublattices are connected by a $C_{2z}$ rotation combined with a fractional translation T =$(0,0,\tfrac{1}{2})$ along the $(001)$ direction. Hence, the remaining coset elements of the point-group can be generated by
$[C_2\parallel C_{2z}]$. Together, these define the nontrivial spin Laue group  $R_{\mathrm{s}}^{III}$, expressed as
${}^{2}m{}^{2}m{}^{1}m =[E\parallel 2/m] + [C_{2}\parallel C_{2z}][E\parallel 2/m]$, with characteristic spin-group integer 2. Little-group analysis predicts that along $k_y=0$ and $k_z=0$, spin degeneracy is 
protected by $C_{2y}$ and $C_{2z}$, respectively, while along $k_x=0$ no such protection 
exists, permitting spin splitting. This is confirmed by the band-structure calculations shown in Fig.~\ref{fig:AAAA_type}(c).

The effective Hamiltonian at the $\Gamma$ point, constrained by the A-AFM symmetries, allows only the quadratic invariant $s_zk_yk_z$, as determined from Tables~\ref{tab:sz_transform} and \ref{tab:A-TYPE}. This reproduces the plane-dependent splitting and confirms consistency between 
symmetry classification, the effective model, and DFT results.

\subsubsection{\textbf{G-Type AFM Symmetry}}

\begin{figure*}
    \centering
    \includegraphics[
        width=1.0\linewidth]{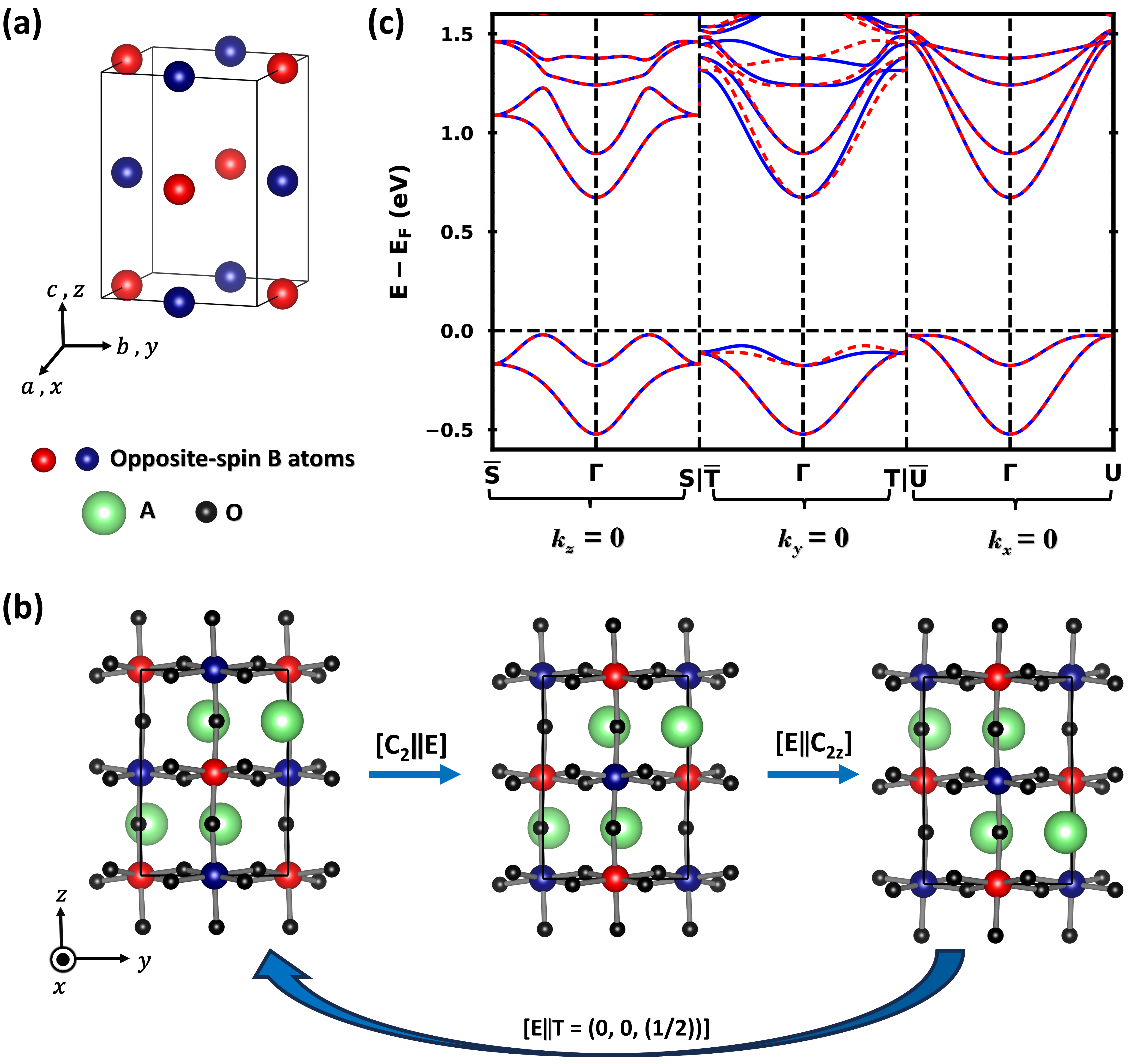}
    \caption{ (a) G-type AFM spin configuration. (b) Symmetry relation between opposite-spin sublattices mediated by a spin-space $C_2$ rotation together with a real-space $C_{2z}$ screw operation and fractional translation $\mathbf{T}=(0,0,\frac12)$. (c) Spin-polarized DFT band structure of G-type AFM LaTiO$_3$ without SOC. Spin splitting appears only in the symmetry-allowed $k_y=0$ plane, while the remaining planes remain spin degenerate. }
    \label{fig:G_type}
\end{figure*}

In G-type AFM (G-AFM), the spins are antiferromagnetically aligned within the $ab$ plane as well as along the $c$ axis (Fig.~\ref{fig:G_type}(a)). 
From the spin point-group consideration, the system belongs to the same ${}^{2}m{}^{2}m{}^{1}m$~\cite{Smejkal2022PRXX} as the C- and A-AFM cases. The halving subgroup consists of $\{E, \bar{E}, m_y, C_{2y}\}$, while the coset generator is $C_{2z}$. As illustrated in Fig.~\ref{fig:G_type}(b), opposite-spin sublattices are connected by a $C_{2z}$ rotation combined with a fractional translation T =$(0,0,\tfrac{1}{2})$ along the $(001)$ direction. As a result, the $C_{2z}$ operation connects opposite-spin sublattices, and the remaining coset elements of the point-group can be generated by
$[C_2\parallel C_{2z}]$. These symmetries define the spin Laue $R_{\mathrm{s}}^{III}$, expressed as
${}^{2}m{}^{2}m{}^{1}m = [E\parallel 2/m] + [C_{2}\parallel C_{2z}][E\parallel 2/m]$, characterized by spin-group integer~2 \cite{Smejkal2022PRXX}.

\begin{table}[b]
\centering
\caption{Same as Table \ref{tab:c-afm-spin}, but for the G-type AFM phase.}
\label{tab:g-TYPE}

\small
\renewcommand{\arraystretch}{1.25}

\resizebox{\columnwidth}{!}{%
\begin{tabular}{|c|c|c|c|c|}
\hline
\textbf{Irreducible Tensor} &
\textbf{$[E\Vert E]$} &
\textbf{$[E\Vert \bar{E}]$} &
\textbf{$[E\Vert C_{2y}]$} &
\textbf{$[C_{2}\Vert C_{2x}]$} \\
\hline

$k_x^2,\;k_y^2,\;k_z^2$
& 1 & 1 & 1 & 1 \\
\hline

$k_xk_y$
& 1 & 1 & $-1$ & $-1$ \\
\hline

$k_yk_z$
& 1 & 1 & $-1$ & 1 \\
\hline

$k_zk_x$
& 1 & 1 & 1 & $-1$ \\
\hline

\end{tabular}
}
\end{table}

Spin degeneracy persists along $k_x=0$ and $k_z=0$, enforced by $C_{2x}$ and $C_{2z}$. Along $k_y=0$, this protection is absent, and spin splitting occurs. Our simulated band structure, as shown in Fig. \ref{fig:G_type}(c),  verify this prediction. The G-AFM symmetry restricts the Hamiltonian to $s_z k_x k_z$, consistent with the observed splitting (Tables~\ref{tab:sz_transform} and \ref{tab:g-TYPE}).

\begin{table*}[t]
\centering
\caption{Calculated structural, magnetic, and electronic properties of the orthorhombic altermagnetic perovskite oxides investigated in this work. For each compound, the space group, magnetic order, N\'eel temperature, experimental and calculated magnetic moments, experimental and calculated band gaps (including the gap character where available), and the symmetry-allowed altermagnetic spin-splitting wavevector plane (ASSWP) are listed. A dash indicates unavailable experimental data.}
\label{tab:allmaterials}

\small
\renewcommand{\arraystretch}{1.25}

\resizebox{\textwidth}{!}{%
\begin{tabular}{|c|l|l|l|c|c|c|c|c|l|c|}
\hline
\textbf{Sr.} &
\textbf{Compound} &
\textbf{Space} &
\textbf{Magnetic} &
\textbf{$T_N$} &
\textbf{$\mu$ (expt.)} &
\textbf{$\mu$ (cal.)} &
\textbf{Band gap} &
\textbf{Band gap} &
\textbf{Band} &
\textbf{ASSWP} \\
\textbf{No.} &
&
\textbf{group} &
\textbf{order} &
\textbf{(K)} &
\textbf{($\mu_B$)} &
\textbf{($\mu_B$)} &
\textbf{(expt.) eV} &
\textbf{(cal.) eV} &
\textbf{nature} &
\textbf{$k_i=0$} \\
\hline

1 & YCrO$_3$  & $Pbnm$~\cite{Judin1966} & G~\cite{Jara2018} & 141~\cite{Judin1966} & 4.5~\cite{Jara2018} & 2.87 & 3.72~\cite{Tiwari2013} & 3.454 & Direct~\cite{Jara2018} & $k_y$ \\
\hline

2 & LaTiO$_3$ & $Pbnm$~\cite{Jin2013} & G~\cite{Jin2013} & 146~\cite{Komarek2007} & 0.57~\cite{Cwik2003} & 0.771 & 0.2~\cite{Jin2013} & 0.625 & Indirect~\cite{Jin2013} & $k_y$ \\
\hline

3 & CaCrO$_3$ & $Pbnm$~\cite{AlarioFranco2009} & C~\cite{Ofer2010_CaCrO3_muSR} & 90~\cite{Goodenough1968_CaCrO3} & 0.295~\cite{Goodenough1968_CaCrO3} & 0.25 & 0~\cite{Weiher1971_CaCrO3} & 0 & Metallic~\cite{Weiher1971_CaCrO3} & $k_z$ \\
\hline

4 & SeNiO$_3$ & $Pnma$~\cite{Munoz2006} & G~\cite{Munoz2006} & 104~\cite{Munoz2006} & 2.11~\cite{Munoz2006} & 1.856 & -- & 3.849 & -- & $k_z$ \\
\hline

5 & SeMnO$_3$ & $Pnma$~\cite{Munoz2006} & G~\cite{Munoz2006} & 53.3~\cite{Munoz2006} & 4.64~\cite{Munoz2006} & 4.659 & -- & 3.779 & -- & $k_z$ \\
\hline

6 & LaFeO$_3$ & $Pbnm$~\cite{Koehler1957} & G~\cite{Koehler1957} & 750~\cite{Koehler1957} & 4.6$\pm$0.2~\cite{Koehler1957} & 4.004 & 2.4~\cite{Koehler1957} & 1.829 & Direct~\cite{Scafetta2014} & $k_y$ \\
\hline

7 & LaCrO$_3$ & $Pbnm$~\cite{Koehler1957} & G~\cite{Koehler1957} & 320~\cite{Koehler1957} & 2.8$\pm$0.2~\cite{Koehler1957} & 2.784 & 2.63~\cite{Koehler1957} & 2.8 & Direct~\cite{Paramanik2018} & $k_y$ \\
\hline

8 & LaMnO$_3$ & $Pbnm$~\cite{Koehler1957} & A~\cite{Koehler1957} & 100~\cite{Koehler1957} & 3.9$\pm$0.2~\cite{Koehler1957} & 3.72 & 1.20~\cite{Koehler1957} & 0.947 & Direct~\cite{Gong2011} & $k_x$ \\
\hline

\end{tabular}%
}
\end{table*}

\subsection{Magnetic symmetry with spin–orbit coupling}

When spin--orbit coupling (SOC) is included, spin orientations get locked to the crystal lattice, 
lowering the symmetry and requiring a shift from the spin space group to a magnetic space group. In orthorhombic perovskites with $Pbnm$ symmetry, the transition-metal ions occupy four equivalent sites in the primitive cell, and once a spin orientation is specified along the crystallographic axes, the resulting ferromagnetic ($F$) and $A$-, $C$-, and $G$-type antiferromagnetic modes at the Brillouin-zone center ($\mathbf{k}=0$) fall into four one-dimensional irreducible representations ($\Gamma_{1}$--$\Gamma_{4}$)$~\cite{Bertaut1968}$.
Non-symmorphic operations such as screw rotations and glide planes map between sublattices, causing 
certain FM and AFM orders to transform identically within a single 
representation~\cite{NonsymmorphicSymmetry}. As a result, weak FM canting or small orbital contributions can appear in 
an AFM phase without additional symmetry reduction, and anomalous Hall conductivity 
(AHC) becomes symmetry-allowed even in nearly compensated states~\cite{WeakFM_AHE}. The degree of canting between two neighboring spins $S_i$ and $S_j$ is determined by the balance between the isotropic Heisenberg exchange $J_{ij}$, which sets the preferred parallel or antiparallel alignment of the spins, and the Dzyaloshinskii--Moriya vector $\mathbf{D}_{ij}$, which favors perpendicular alignment between them$~\cite{DM}$. Single-ion anisotropy (SIA) further determines the preferred orientation of the spins relative to the crystal axes and can favor specific canting patterns depending on lattice symmetry~\cite{SIAA}.
% \textcolor{green}{Single-ion anisotropy (SIA), on the other hand, pins the spin orientations relative to the crystal axes and can favor specific canting patterns depending on lattice symmetry~\cite{SIAA}. Together, these interactions $J_{ij}$, $\mathbf{D}_{ij}$, and SIA control the resulting weak FM or weak AFM components, with the canting angle $\theta_{ij}$ approximately given by~\cite{CantingGG}
% \[
% \theta_{ij} \sim \arctan\left(\frac{|\mathbf{D}_{ij}|}{|J_{ij}|}\right).
% \]}

More generally, when two of the 
three key symmetries (mirror $\perp z$, b-glide $\perp x$, n-glide $\perp y$) are broken by an 
AFM order, the resulting Berry curvature acts as an effective magnetic field, producing 
a transverse Hall response in the plane perpendicular to the intersection of the broken symmetries~\cite{Vanderbilt2018}.
Only when all three are preserved does the anomalous Hall effect vanish, linking the perovskite case 
to the broader family of altermagnets~\cite{altermagnetismFound}.

To microscopically trace how these broken symmetries give rise to $k$-space spin splittings, we construct the effective Hamiltonian at the $\Gamma$ point. Specifically, we examine all second-order irreducible momentum tensors, $k_x^2, k_y^2, k_z^2, k_x k_y, k_x k_z, k_y k_z$, with respect to the magnetic space-group symmetries. Since spin-orbit coupling (SOC) is present, we analyze how each spin component ($s_x, s_y, s_z$) transforms under the little group of each momentum tensor, identifying which spin components remain invariant (i.e., transform as the identity representation). Spin components that are invariant under all operations of the little group are symmetry-allowed and give rise to spin splittings, corresponding to ferromagnetic-like contributions and thereby symmetry-permitted spin-polarized terms in the Hamiltonian.

Table~\ref{tab:allmaterials} shows the space group, magnetic structure,  N\'eel temperature, calculated and experimental magnetic moments and band gaps, and their nature for a list of orthorhombic altermagnetic perovskite oxides.
Although the symmetry and Hamiltonian analysis applies to all these materials, we show the band structure  only for LaTiO$_3$ and CaCrO$_3$. This choice is motivated by the subsequent calculation of their AHC (as LaTiO$_3$ is a Mott insulator with the smallest band gap whereas CaCrO$_3$ is metallic, as summarised in Table~\ref{tab:allmaterials}) in Sec. III(D), which allows us to directly connect the symmetry-constrained spin polarizations to their AHC.

\begin{figure*}
    \centering
    \includegraphics[
        width=1.0\linewidth]{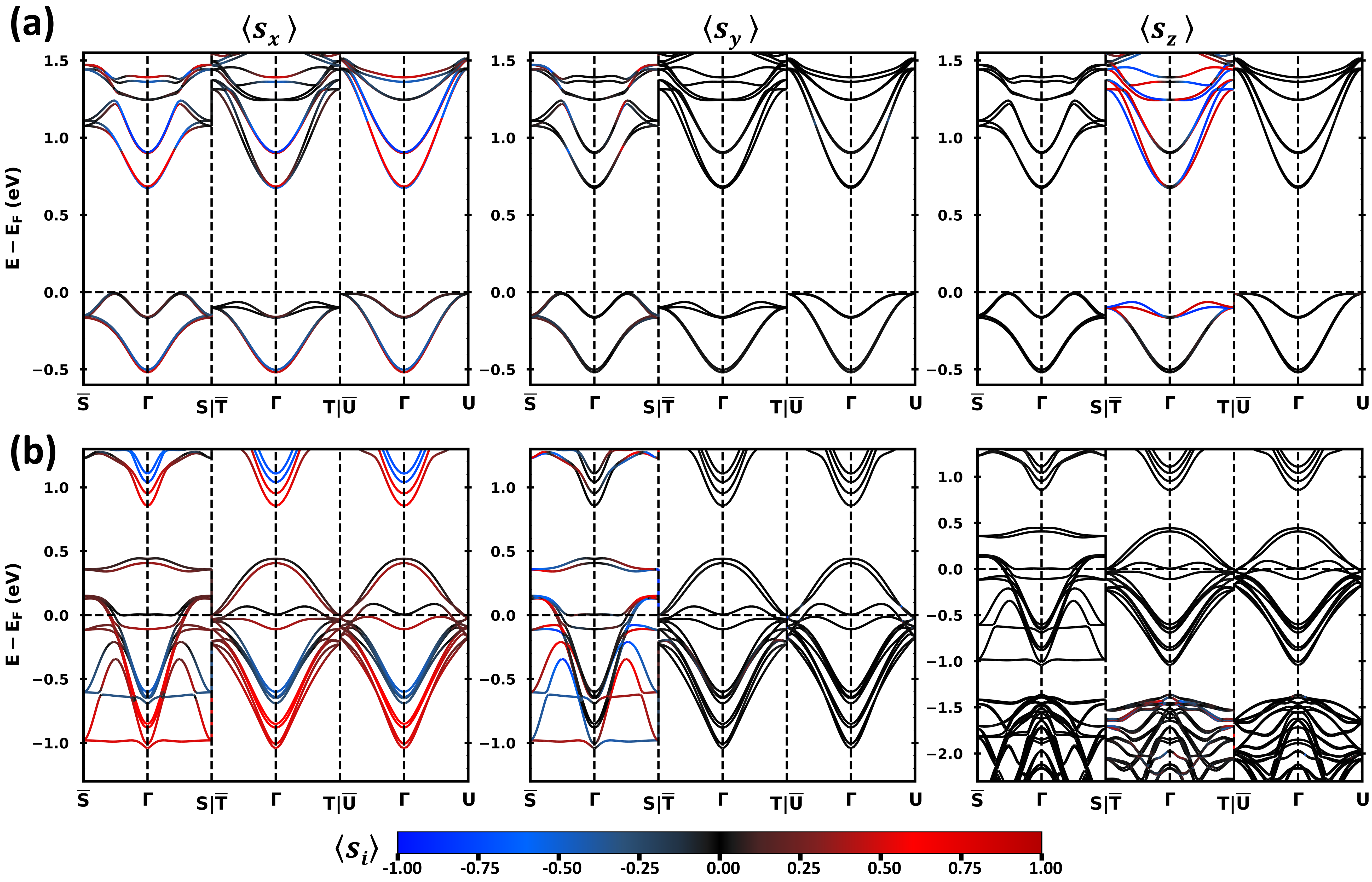}
    \caption{Spin-resolved DFT band structures including SOC for the $\Gamma_2$ magnetic configuration in (a) LaTiO$_3$ ($F_xC_y\bm{G_z}$) and (b) CaCrO$_3$ ($F_x\bm{C_y}G_z$). Here, $\langle s_i \rangle$ ($i = x, y, z$) represents the expected value of spin polarization projected along the $x, y,$ and $z$ directions, as indicated by the color scale. According to the symmetry analysis, the additional SOC-induced spin splitting is expected to be associated with the symmetry-allowed $s_x$ spin polarization arising from weak FM canting while preserving the nonrelativistic altermagnetic spin splitting. cd A  
    }
    \label{fig:FxCyGz}
\end{figure*}

\begin{figure*}
    \centering
    \includegraphics[
        width=1.0\linewidth]{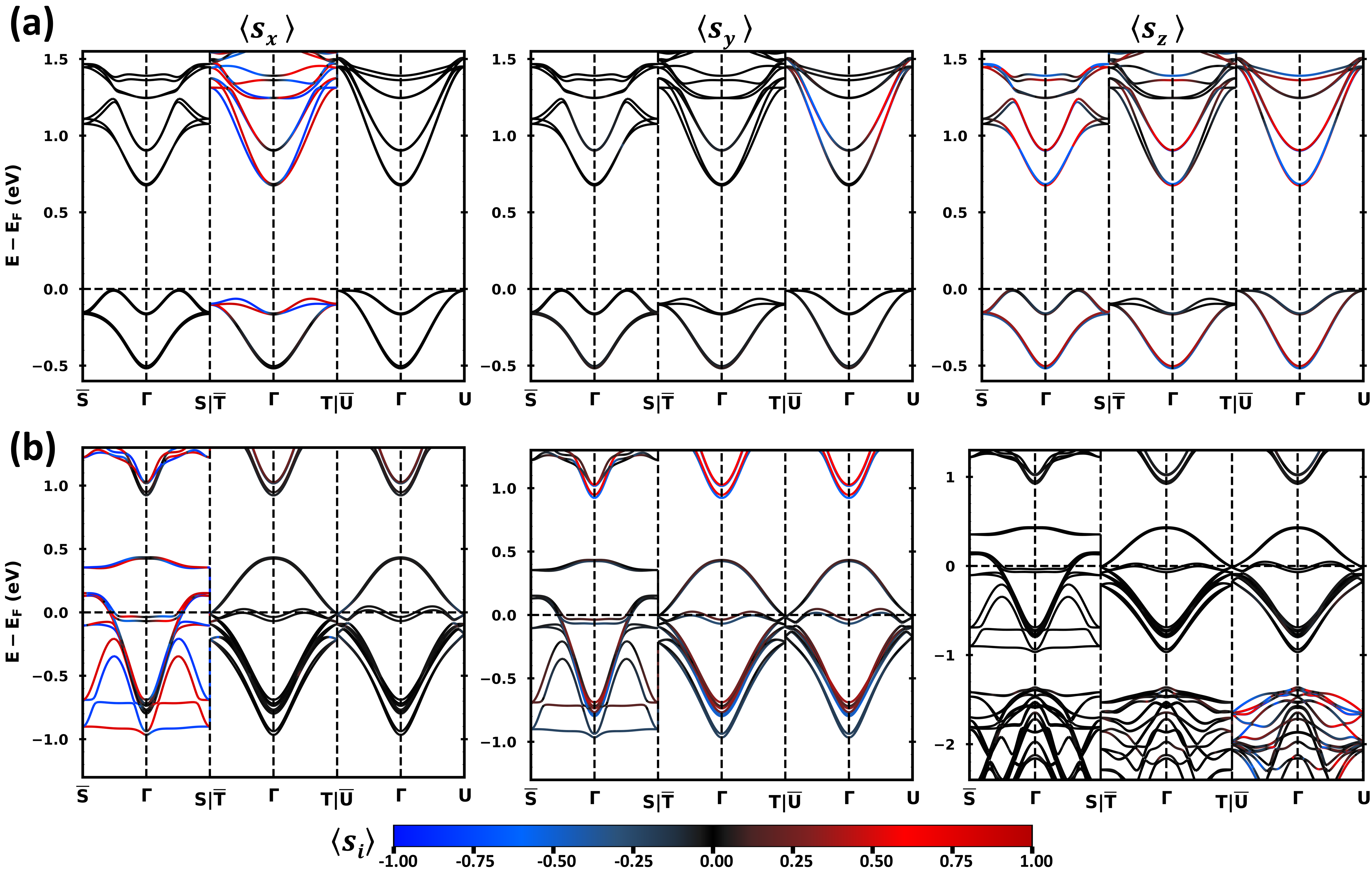}
    \caption{ Spin-resolved DFT band structures including SOC for the (a) $\Gamma_3$ ($\bm{G_x} A_y F_z$) configuration of LaTiO$_3$ and (b) the $\Gamma_1$ ($\bm{C_x} A_z F_y$) configuration of CaCrO$_3$. The SOC-induced spin splitting is expected to be associated with the symmetry-allowed weak FM components corresponding to the respective irreducible representations.}
    \label{fig:gamma1nd3}
\end{figure*}

\begin{figure*}
    \centering
    \includegraphics[
        width=1.0\linewidth]{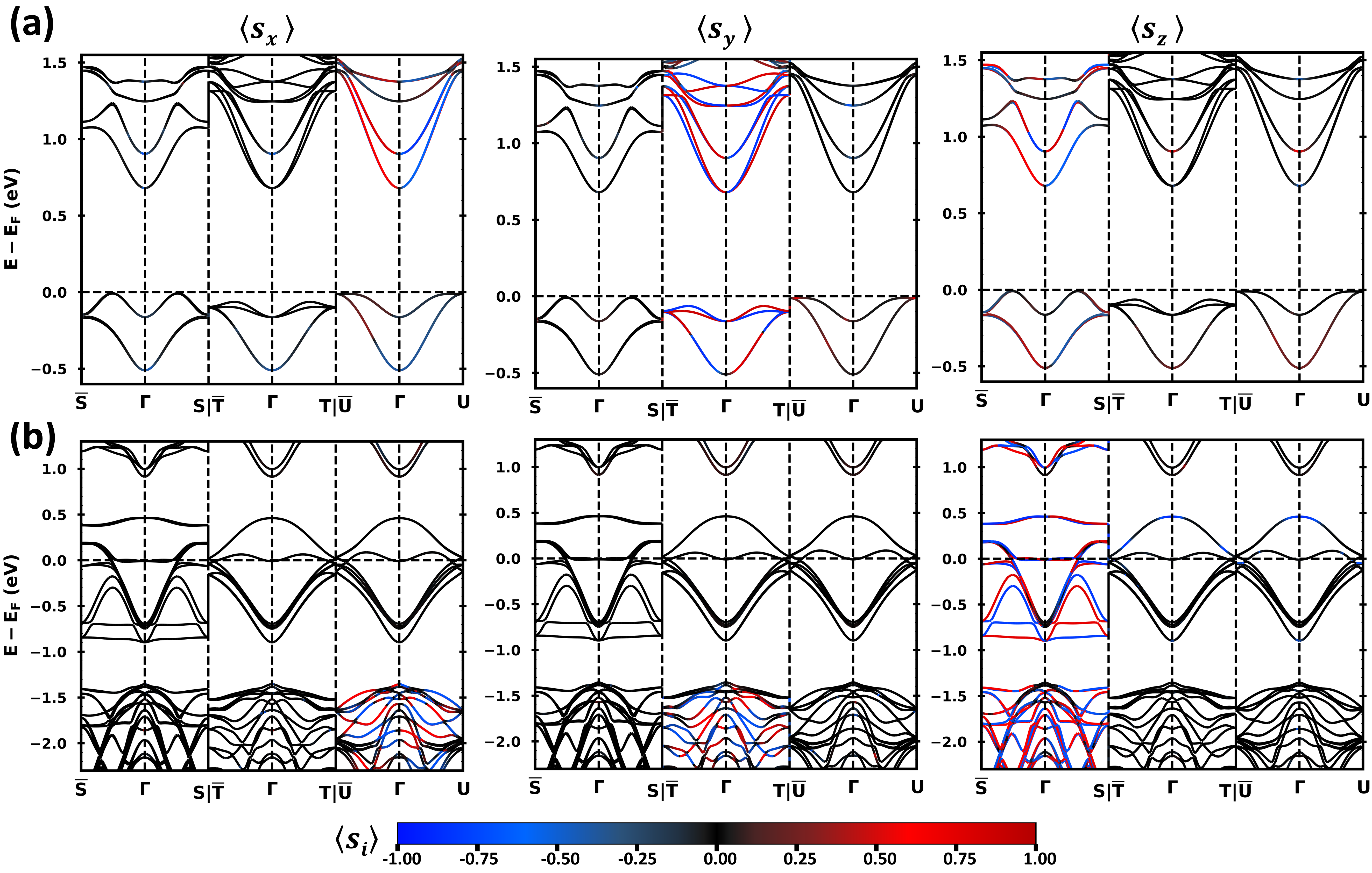}
        \caption{Spin-resolved DFT band structures including SOC for the $\Gamma_4$ magnetic configuration in (a) LaTiO$_3$ ($A_x\bm{G_y}C_z$) and (b) CaCrO$_3$ ($A_xG_y\bm{C_z}$). Since weak FM canting is symmetry forbidden in $\Gamma_4$, the observed spin splitting originates solely from the collinear altermagnetic order and closely resembles the non-relativistic band structure.}
    \label{fig:gamma4}
\end{figure*}

Figures ~\ref{fig:FxCyGz} --~\ref{fig:gamma4} show the simulated band structures of LaTiO$_3$ and CaCrO$_3$ including the effect of SOC with the spin polarization ($\langle s_i\rangle$) highlighted by the color bar. The captions specify the magnetic orders in Bertaut’s notation, with collinear ones indicated in \textbf{bold}. We will now explain the symmetry of each of these magnetic configurations for these two systems in some detail below.

\begin{table*}[t]
\centering
\caption{Symmetry classification of collinear antiferromagnetic orders in orthorhombic $Pbnm$ perovskites. For each irreducible representation ($\Gamma_1$--$\Gamma_4$), the corresponding magnetic configuration, symmetry generators, preserved nonsymmorphic symmetry, symmetry-allowed weak ferromagnetic moment, allowed anomalous Hall conductivity tensor component, and magnetic space group are listed. Here, ``$+$'' and ``$-$'' indicate the characters of symmetry operations, with ``$+$'' assigned to elements and ``$-$'' to antielements, following the magnetic representation formalism~\cite{Bertaut1968}.}
\label{tab:magnetic_irreps}

\small
\renewcommand{\arraystretch}{1.25}

\begin{tabular}{|c|c|c|c|c|c|}
\hline
\textbf{Irrep (AFM order)} &
\textbf{Generators ($\bar{E}, C_{2x}, C_{2y}$)} &
\textbf{Preserved symmetry} &
\textbf{Induced FM} &
\textbf{AHE tensor} &
\textbf{Magnetic SG} \\
\hline

$\Gamma_1 \equiv C_xF_yA_z$ &
$(+,-,+)$ &
$n$-glide ($\perp y$) &
$m_y$ &
$\sigma_{xz}$ &
$Pb'nm'$ or $Pnm'a'$ \\
\hline

$\Gamma_2 \equiv F_xC_yG_z$ &
$(+,+,-)$ &
$b$-glide ($\perp x$) &
$m_x$ &
$\sigma_{yz}$ &
$Pbn'm'$ or $Pn'm'a$ \\
\hline

$\Gamma_3 \equiv G_xA_yF_z$ &
$(+,-,-)$ &
Mirror ($\perp z$) &
$m_z$ &
$\sigma_{zx}$ &
$Pb'n'm$ or $Pn'ma'$ \\
\hline

$\Gamma_4 \equiv A_xG_yC_z$ &
$(+,+,+)$ &
All &
None &
Forbidden &
$Pbnm$ or $Pnma$ \\
\hline

\end{tabular}
\end{table*}

\begin{table}[t]
\centering
\caption{Transformation properties of the spin components under the symmetry operations of the $\Gamma_2 \equiv F_xC_yG_z$ magnetic configuration with magnetic space group $Pb'nm'$ in the presence of SOC.}
\label{tab:symmetry_spin_1}

\small
\renewcommand{\arraystretch}{1.25}

\begin{tabular*}{\columnwidth}{@{\extracolsep{\fill}}|c|c|c|c|c|c|c|c|c|}
\hline

\textbf{Spin component} &
\textbf{$E$} &
\textbf{$\bar{E}$} &
\textbf{$C_{2x}$} &
\textbf{$m_x$} &
\textbf{$\tau C_{2y}$} &
\textbf{$\tau C_{2z}$} &
\textbf{$\tau m_y$} &
\textbf{$\tau m_z$} \\
\hline

$s_x$ &
1 & 1 & 1 & 1 & 1 & 1 & 1 & 1 \\
\hline

$s_y$ &
1 & 1 & $-1$ & $-1$ & $-1$ & 1 & $-1$ & 1 \\
\hline

$s_z$ &
1 & 1 & $-1$ & $-1$ & 1 & $-1$ & 1 & $-1$ \\
\hline

\end{tabular*}

\end{table}

\begin{table*}[t]
\centering
\caption{Symmetry-allowed spin components and the corresponding Bertaut magnetic modes for the off-diagonal irreducible momentum tensors up to second order in the $\Gamma_2 \equiv F_xC_yG_z$ irreducible representation. The diagonal tensors $k_x^2$, $k_y^2$, and $k_z^2$ are omitted since they remain invariant under all symmetry operations and therefore impose no additional symmetry constraints.}
\label{tab:irrep_tensor}

\small
\renewcommand{\arraystretch}{1.25}

\begin{tabular*}{\textwidth}{@{\extracolsep{\fill}}|l|c|c|c|}
\hline

\textbf{Irreducible tensor} &
\textbf{Surviving little group} &
\textbf{Allowed spin components} &
\textbf{Symmetry-allowed magnetic modes} \\
\hline

$k_xk_y$ &
$\{E,\;\bar{E},\;\tau C_{2z},\;\tau m_z\}$ &
$s_x,\; s_y$ &
$F_x,\; C_y$ \\
\hline

$k_yk_z$ &
$\{E,\;\bar{E},\;C_{2x},\;m_x\}$ &
$s_x$ &
$F_x$ \\
\hline

$k_zk_x$ &
$\{E,\;\bar{E},\;\tau C_{2y},\;\tau m_y\}$ &
$s_x,\; s_z$ &
$F_x,\; G_z$ \\
\hline

\end{tabular*}

\end{table*}

\begin{table}[b]
\centering
\caption{Same as Table~\ref{tab:symmetry_spin_1}, but for the $\Gamma_1 \equiv C_xF_yA_z$ irreducible representation.}
\label{tab:symmetry_spin_2}

\small
\renewcommand{\arraystretch}{1.25}

\begin{tabular*}{\columnwidth}{@{\extracolsep{\fill}}|c|c|c|c|c|c|c|c|c|}
\hline

\textbf{Spin component} &
\textbf{$E$} &
\textbf{$\bar{E}$} &
\textbf{$C_{2y}$} &
\textbf{$m_y$} &
\textbf{$\tau C_{2z}$} &
\textbf{$\tau C_{2x}$} &
\textbf{$\tau m_z$} &
\textbf{$\tau m_x$} \\
\hline

$s_x$ &
1 & 1 & $-1$ & $-1$ & 1 & $-1$ & 1 & $-1$ \\
\hline

$s_y$ &
1 & 1 & 1 & 1 & 1 & 1 & 1 & 1 \\
\hline

$s_z$ &
1 & 1 & $-1$ & $-1$ & $-1$ & 1 & $-1$ & 1 \\
\hline

\end{tabular*}

\end{table}

\begin{table}[b]
\centering
\caption{Same as Table~\ref{tab:symmetry_spin_1}, but for the $\Gamma_3 \equiv G_xA_yF_z$ irreducible representation.}
\label{tab:symmetry_spin_3}

\small
\renewcommand{\arraystretch}{1.25}

\begin{tabular*}{\columnwidth}{@{\extracolsep{\fill}}|c|c|c|c|c|c|c|c|c|}
\hline

\textbf{Spin component} &
\textbf{$E$} &
\textbf{$\bar{E}$} &
\textbf{$C_{2z}$} &
\textbf{$m_z$} &
\textbf{$\tau C_{2y}$} &
\textbf{$\tau C_{2x}$} &
\textbf{$\tau m_y$} &
\textbf{$\tau m_x$} \\
\hline

$s_x$ &
1 & 1 & $-1$ & $-1$ & 1 & $-1$ & 1 & $-1$ \\
\hline

$s_y$ &
1 & 1 & $-1$ & $-1$ & $-1$ & 1 & $-1$ & 1 \\
\hline

$s_z$ &
1 & 1 & 1 & 1 & 1 & 1 & 1 & 1 \\
\hline

\end{tabular*}

\end{table}

\begin{table*}
\centering
\caption{Same as Table~\ref{tab:irrep_tensor}, but for the $\Gamma_1 \equiv C_xF_yA_z$ irreducible representation.}
\label{tab:irrep_tensor_1}

\small
\renewcommand{\arraystretch}{1.25}

\begin{tabular*}{\textwidth}{@{\extracolsep{\fill}}|l|c|c|c|}
\hline

\textbf{Irreducible tensor} &
\textbf{Surviving little group} &
\textbf{Allowed spin components} &
\textbf{Symmetry-allowed magnetic modes} \\
\hline

$k_xk_y$ &
$\{E,\;\bar{E},\;\tau C_{2z},\;\tau m_z\}$ &
$s_y,\;s_x$ &
$F_y,\; C_x$ \\
\hline

$k_yk_z$ &
$\{E,\;\bar{E},\;\tau C_{2x},\;\tau m_x\}$ &
$s_y,\;s_z$ &
$F_y,\; A_z$ \\
\hline

$k_zk_x$ &
$\{E,\;\bar{E}, C_{2y},  m_y\}$ &
$s_y$ &
$F_y$ \\
\hline

\end{tabular*}

\end{table*}

\begin{table*}
\centering
\caption{Same as Table~\ref{tab:irrep_tensor}, but for the $\Gamma_3 \equiv G_xA_yF_z$ irreducible representation.}
\label{tab:irrep_tensor_2}

\small
\renewcommand{\arraystretch}{1.25}

\begin{tabular*}{\textwidth}{@{\extracolsep{\fill}}|l|c|c|c|}
\hline

\textbf{Irreducible tensor} &
\textbf{Surviving little group} &
\textbf{Allowed spin components} &
\textbf{Symmetry-allowed magnetic modes} \\
\hline

$k_xk_y$ &
$\{E,\;\bar{E},\; C_{2z},\; m_z\}$ &
$s_z$ &
$F_z$ \\
\hline

$k_yk_z$ &
$\{E,\;\bar{E},\;\tau C_{2x},\;\tau m_x\}$ &
$s_z,\;s_y$ &
$F_z,\; A_y$ \\
\hline

$k_zk_x$ &
$\{E,\;\bar{E},\;\tau C_{2y},\;\tau m_y\}$ &
$s_z,\; s_x$ &
$F_z,\; G_x$ \\
\hline

\end{tabular*}

\end{table*}

% \subsubsection{\textbf{\boldmath{\Gamma_2 \equiv F_x C_y G_z$ }}}
\subsubsection{$\bm{\Gamma_2 \equiv F_x C_y G_z}$}

In the $F_xC_yG_z$ configuration, the three modes $F_x$, $C_y$, and $G_z$ belong to the same irreducible representation. 
This implies that if the system is in a collinear C-type AFM state with spins aligned along the global $y$ axis, symmetry allows a weak FM spin canting along $x$ and a weak G-type AFM spin canting along $z$. Conversely, if the primary order is G-AFM (spins along $z$), symmetry allows a weak FM spin canting along $x$ and a weak C-type AFM spin canting along $y$. The corresponding magnetic space group is $Pbn'm'$ or $Pn'm'a$~\cite{Bertaut1968}. The corresponding magnetic point group, $m'm'm$, comprises eight symmetry operations and is derivable from three generators: two unitary symmetries $\bar{E}$ and $C_{2x}$, one antiunitary symmetry, $\tau C_{2y}$.

With the magnetic symmetry of the $\Gamma_2$ phase established, we next examine how these symmetry operations constrain the allowed spin polarizations and determine the effective Hamiltonian at the $\Gamma$ point. To this end, we consider all second-order irreducible momentum tensors. Since $k_x^2$, $k_y^2$, and $k_z^2$ remain invariant under all symmetry operations, and $s_x$ is likewise invariant under every operation of the magnetic space group (Table~\ref{tab:symmetry_spin_1}), an $s_x$-polarized contribution giving rise to $F_x$ canting is symmetry-allowed throughout the Brillouin zone. The remaining off-diagonal momentum tensors determine the additional symmetry-allowed spin components on different momentum planes as summarized in Table~\ref{tab:irrep_tensor}. These symmetry constraints are naturally incorporated into the effective Hamiltonian, providing a compact framework for predicting the symmetry-allowed spin polarizations associated with different momentum tensors. The corresponding effective
Hamiltonian is, therefore, given by Eq.~\eqref{eq:11},
where $A$, $B$, and $C$ denote real coefficients.

\begin{align}
\label{eq:11}
H_{\Gamma_2} =
A s_{x}\left(k_{x}^{2} + k_{y}^{2} + k_{z}^{2}\right)
+ k_{x}k_{z}\left(As_{x} + Cs_{z}\right) \notag \\
+ k_{x}k_{y}\left(As_{x} + Bs_{y}\right)
+ k_{y}k_{z}\left(As_{x}\right).
\end{align}

Specifically, the isotropic term $A(k_x^2+k_y^2+k_z^2)s_x$ guarantees the presence of an $s_x$ component throughout the Brillouin zone. The $k_xk_y(As_x+Bs_y)$ term shows that the $k_z=0$ plane supports both $s_x$ and $s_y$ spin polarizations, whereas the $k_xk_z(As_x+Cs_z)$ term permits $s_x$ and $s_z$ on the $k_y=0$ plane. In contrast, the $k_yk_z(As_x)$ term contains only the $s_x$ contribution, implying that only the $s_x$ component survives on the $k_x=0$ plane. 

This behavior mirrors the nonrelativistic analysis: in C-AFM order, spin polarization is enforced on the $k_z=0$ plane, whereas in G-AFM order it occurs on the $k_y=0$ plane. The SOC therefore ensures that $s_x$ polarization arises universally, $s_y$ polarization originates from the C-type AFM channel, and $s_z$ polarization is induced by the G-type AFM contribution.

Finally, we validate our group-theoretical predictions against the calculated spin-resolved SOC band structures. As shown in Fig.~\ref{fig:FxCyGz}(a), SOC induces additional spin splitting between 0.8 and 1.0~eV in LaTiO$_3$ ($G_z$) compared to the nonrelativistic limit in Fig.~\ref{fig:G_type}(c). Likewise, SOC gives rise to notable spin splittings between $-1.0$ and 0.5~eV in CaCrO$_3$ ($C_y$) (Fig.~\ref{fig:FxCyGz}(b)) relative to the nonrelativistic band structure in Fig.~\ref{fig:CCC-AFMMM}(d).

Crucially, the spin projections $\langle s_i \rangle$ ($i = x, y, z$) obtained from first-principles calculations (Fig.~\ref{fig:FxCyGz}) explicitly confirm the symmetry constraints dictated by Eq.~\eqref{eq:11}. Because the $\Gamma_2$ irreducible representation permits the $F_x$ magnetic mode, an $s_x$ spin polarization emerges universally across all high-symmetry momentum planes ($k_x = 0$, $k_y = 0$, and $k_z = 0$) for both materials shown in Fig.~\ref{fig:FxCyGz}.

In LaTiO$_3$, where the primary order parameter is G-type AFM ($G_z$), a pronounced, momentum-dependent $s_z$ spin polarization develops on the $k_y = 0$ plane. Because the underlying $G_z$ order is antiferromagnetic, this $s_z$ polarization is inherently momentum-compensated, exhibiting alternating spin signs across $k$-space. In contrast, the symmetry-allowed $C_y$ canting in LaTiO$_3$ is extremely weak; consequently, the $s_y$ polarization on the $k_z = 0$ plane is nearly vanishing, displaying only faint, compensated signatures around 1.5~eV (Fig.~\ref{fig:FxCyGz}(a)).

Similarly, in CaCrO$_3$, the dominant C-type AFM ordering ($C_y$) enforces a strong $s_y$ spin polarization on the $k_z = 0$ plane, which is likewise momentum-dependent and spin-compensated as dictated by the altermagnetic point-group symmetry. Meanwhile, the $G_z$ mode originating from $z$-axis canting in CaCrO$_3$ is negligible, leaving the $s_z$ polarization on the $k_y = 0$ plane virtually zero, with only faint, compensated features visible at deeper energies around $-1.5$~eV (Fig.~\ref{fig:FxCyGz}(b)).

This remarkable agreement between analytical symmetry operations and first-principles electronic structures establishes the effective Hamiltonian as a quantitative framework for mapping symmetry-enforced spin textures. Beyond validating the microscopic model, these momentum-dependent, compensated spin polarizations provide a concrete physical foundation for controlling spin-polarized currents and transport responses in functional altermagnetic perovskites.

\begin{figure}[t]
    \centering
    \includegraphics[
        width=1.0\linewidth]{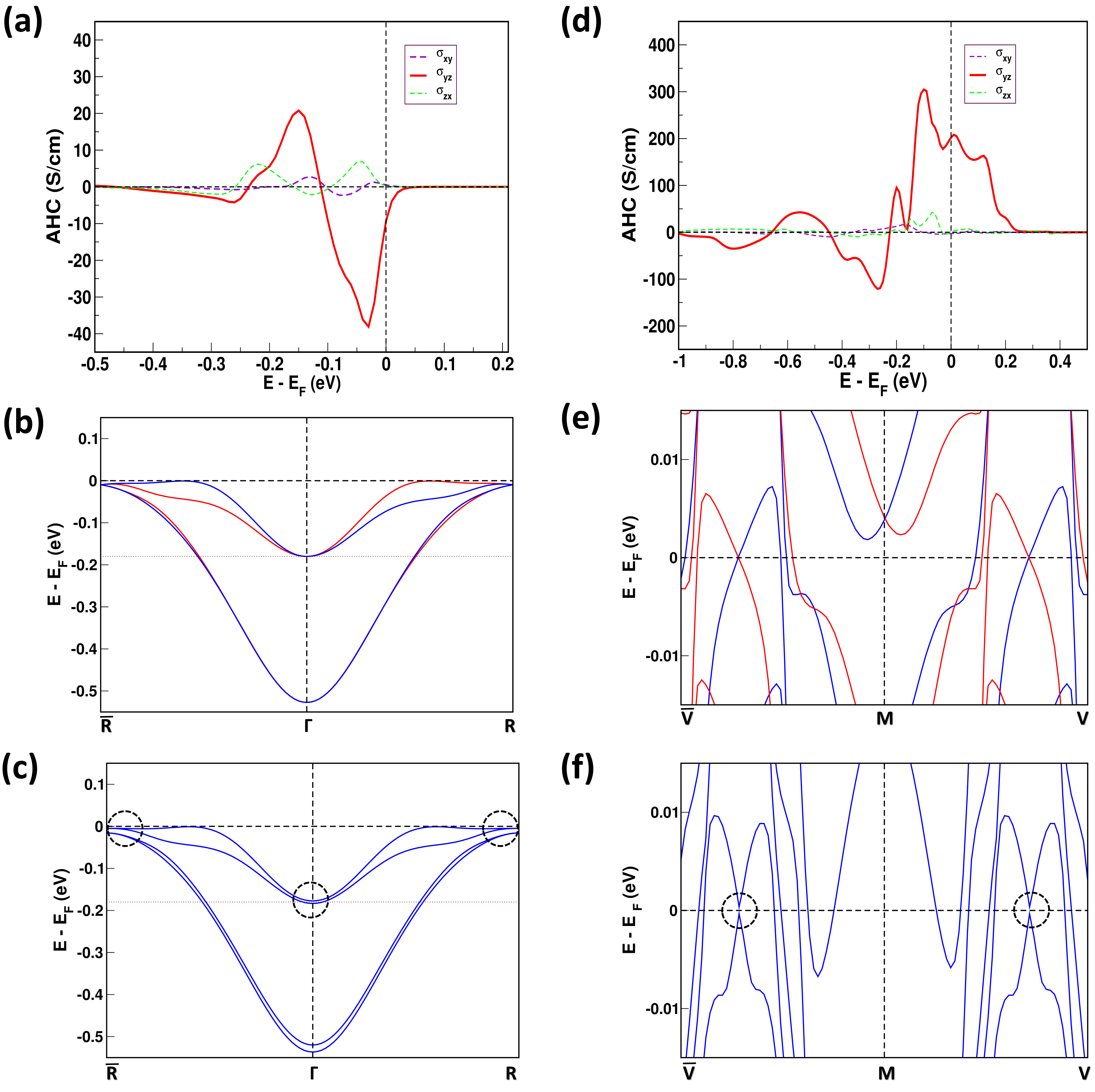}
        \caption{Anomalous Hall conductivity (AHC) and its microscopic origin. (a,d) Energy dependence of the symmetry-allowed AHC tensor components for LaTiO$_3$ ($\bm{G_z}C_yF_x$) and CaCrO$_3$ ($\bm{C_y}G_zF_x$). (b,c) Electronic band structures of LaTiO$_3$ along the high-Berry-curvature path along the $\bar{R}(-0.5,\,0.5,\,0.5)$--$\Gamma(0,\,0,\,0)$--$R(0.5,\,0.5,\,0.5)$, without and with SOC, respectively. (e,f) Corresponding band structures of CaCrO$_3$ along the $\bar{V}(-0.06,\,-0.7,\,0.49)$--$M(0,\,0,\,0)$--$V(-0.06,\,0.7,\,0.49)$. In both materials, spin-polarized band crossings already exist in the absence of SOC due to altermagnetic symmetry. SOC opens small anticrossing gaps at these crossings, generating enhanced Berry curvature and the resulting AHC.}
    \label{fig:ahc}
\end{figure}

% \begin{widetext}
%     1. \textcolor{red}{https://pyprocar.readthedocs.io/en/stable/examples/00-band\_structure/01-Dealing\%20with\%20Spin.html} \\
%     2. \textcolor{red}{https://pyprocar.readthedocs.io/en/stable/user-guide/filter.html} \\
%     3. \textcolor{red}{https://pyprocar.readthedocs.io/en/stable/examples/00-band\_structure/00-Getting\%20Started.html}
% \end{widetext}

% \subsubsection{\textbf{\boldmath{$\Gamma_1 \equiv C_x A_z F_y$ and $\Gamma_3 \equiv G_x A_y F_z$}}}
\subsubsection{$\bm{\Gamma_1 \equiv C_x F_y A_z}$ and $\bm{\Gamma_3 \equiv G_x A_y F_z}$}

As in $\Gamma_2$, each of the configurations $\Gamma_1 \,\equiv C_x A_z F_y$ and $\Gamma_3 \,\equiv C_z A_x F_y$ has three magnetic modes belonging to the same irreducible representation, allowing symmetry-permitted spin canting analogous to that discussed for $\Gamma_2$. The magnetic space group of $\Gamma_1$ ($\Gamma_3$) is $Pb'nm'$ or $Pnm'a'$
($Pb'n'm$ or $Pn'ma'$), with the corresponding magnetic point group
$mm'm'$ ($m'mm'$), generated by $\bar{E},C_{2y},\, \tau C_{2x}$ ($\bar{E},\, \tau C_{2x},\, \tau C_{2y}$).

Following the same procedure as for the $\Gamma_2$ phase, the transformation properties of the spin components summarized in Tables~\ref{tab:symmetry_spin_2} and \ref{tab:symmetry_spin_3} for the $\Gamma_1$ and $\Gamma_3$ phases, respectively, together with the corresponding symmetry-allowed spin components for the irreducible momentum tensors summarized in Tables~\ref{tab:irrep_tensor_1} and \ref{tab:irrep_tensor_2}, determine the effective Hamiltonians at the $\Gamma$ point, given by Eqs.~\eqref{eq:12} and \eqref{eq:13}.

\begin{align}
\label{eq:12}
H_{\Gamma_1} 
= A s_y \left( k_x^2 + k_y^2 + k_z^2 \right) 
+ k_x k_y \left( A s_y + B s_x \right) \notag \\
+ k_y k_z \left( A s_y + C s_z \right) 
+ k_z k_x \left( A s_y \right)
\end{align}

\begin{align}
\label{eq:13}
H_{\Gamma_3}  = 
As_z \left( k_x^2 + k_y^2 + k_z^2 \right) 
+ k_x k_z \left( A s_x + C s_z \right) \notag \\
+ k_y k_z \left( A s_y + B s_z \right) 
+ k_x k_y \left( A s_z \right)
\end{align}

As established for the $\Gamma_2$ phase, the spin-resolved band structures for the $\Gamma_3$ configuration in LaTiO$_3$ ($G_x$) (Fig.~\ref{fig:gamma1nd3}(a)) and the $\Gamma_1$ configuration in CaCrO$_3$ ($C_x$) (Fig.~\ref{fig:gamma1nd3}(b)) exhibit precise agreement with the symmetry rules derived in Eqs.~\eqref{eq:12} and \eqref{eq:13}.

For the $\Gamma_3$ phase in LaTiO$_3$ (Fig.~\ref{fig:gamma1nd3}(a)), the primary $G_x$ magnetic order generates a dominant, momentum-compensated $s_x$ spin polarization on the $k_y = 0$ plane. Simultaneously, the symmetry-permitted $F_z$ mode drives a universal $s_z$ spin polarization across all high-symmetry momentum planes. In contrast, the weak $A_y$ mode results in a nearly vanishing $s_y$ polarization on the $k_x = 0$ plane, displaying only faint, compensated features around 1.0~eV.\\
Similarly, for the $\Gamma_1$ phase in CaCrO$_3$ (Fig.~\ref{fig:gamma1nd3}(b)), the primary $C_x$ ordering dictates a strong, momentum-compensated $s_x$ spin polarization on the $k_z = 0$ plane. The allowed $F_y$ mode ensures a universal $s_y$ spin polarization throughout the Brillouin zone, whereas the weak $A_z$ mode leaves the $s_z$ polarization on the $k_x = 0$ plane virtually zero, with only faint features visible at deeper energies around $-1.5$~eV.

This precise correspondence confirms that the primary magnetic orders ($G_x$ and $C_x$) govern the dominant momentum-compensated $s_x$ spin textures, while the ferromagnetic components ($F_z$ and $F_y$) dictate universal spin polarizations, fully validating the effective Hamiltonians in Eqs.~\eqref{eq:12} and \eqref{eq:13}.

% \subsubsection{\textbf{\boldmath{$\Gamma_4 \equiv A_xG_yC_z$ }}}
\subsubsection{$\bm{\Gamma_4 \equiv A_xG_yC_z}$}

\begin{table}[b]
\centering
\caption{Same as Table~\ref{tab:symmetry_spin_1}, but for the $\Gamma_4 \equiv A_xG_yC_z$ irreducible representation.}
\label{tab:symmetry_spin}

\small
\renewcommand{\arraystretch}{1.25}

\begin{tabular*}{\columnwidth}{@{\extracolsep{\fill}}|c|c|c|c|c|c|c|c|c|}
\hline

\textbf{Spin component} &
\textbf{$E$} &
\textbf{$\bar{E}$} &
\textbf{$C_{2x}$} &
\textbf{$C_{2y}$} &
\textbf{$C_{2z}$} &
\textbf{$m_x$} &
\textbf{$m_y$} &
\textbf{$m_z$} \\
\hline

$s_x$ &
1 & 1 & 1 & $-1$ & $-1$ & 1 & $-1$ & $-1$ \\
\hline

$s_y$ &
1 & 1 & $-1$ & 1 & $-1$ & $-1$ & 1 & $-1$ \\
\hline

$s_z$ &
1 & 1 & $-1$ & $-1$ & 1 & $-1$ & $-1$ & 1 \\
\hline

\end{tabular*}

\end{table}

\begin{table*}[t]
\centering
\caption{Same as Table~\ref{tab:irrep_tensor}, but for the $\Gamma_4 \equiv A_xG_yC_z$ irreducible representation.}
\label{tab:irrep_tensor_4}

\small
\renewcommand{\arraystretch}{1.25}

\begin{tabular*}{\textwidth}{@{\extracolsep{\fill}}|l|c|c|c|}
\hline

\textbf{Irreducible tensor} &
\textbf{Surviving little group} &
\textbf{Allowed spin components} &
\textbf{Symmetry-allowed magnetic modes} \\
\hline

$k_xk_y$ &
$\{E,\;\bar{E}, C_{2z}, m_z\}$ &
$s_z$ &
$C_z$ \\
\hline

$k_yk_z$ &
$\{E,\;\bar{E}, C_{2x}, m_x\}$ &
$s_x$ &
$A_x$ \\
\hline

$k_zk_x$ &
$\{E,\;\bar{E}, C_{2y},  m_y\}$ &
$s_y$ &
$G_y$ \\
\hline

\end{tabular*}

\end{table*}

In the $\Gamma_4$ configuration, the three modes $A_x$, $G_y$, and $C_z$ belong to the same irreducible representation. The magnetic space group is $Pbnm$, identical to the parent space group,
with point group $mmm$, generated by $\bar{E}$, $C_{2x}$, and $C_{2y}$. Unlike the previously discussed irreducible representations ($\Gamma_1$–$\Gamma_3$), FM canting is symmetry-forbidden, so spin splitting arises solely from the AFM order parameters. Examining all second-order irreducible momentum tensors against the magnetic space-group operations reveals that while $k_x^2$, $k_y^2$, and $k_z^2$ remain invariant, no spin component remains invariant under any operations other than $E$ and $\bar{E}$.

Following the same methodology applied to the $\Gamma_1$–$\Gamma_3$ phases, the symmetry constraints summarized in Tables~\ref{tab:symmetry_spin} and \ref{tab:irrep_tensor_4} determine the effective Hamiltonian at the $\Gamma$ point, given by Eq.~\eqref{eq:4gamma}.

\begin{align}
\label{eq:4gamma}
H_{\Gamma_4} = A s_x \, k_y k_z + B s_y \, k_z k_x + C s_z \, k_x k_y
\end{align}

The effective Hamiltonian in Eq.~\eqref{eq:4gamma} highlights the absence of any isotropic spin term. Instead, the momentum-dependent spin polarizations are strictly plane-selective: on the $k_x = 0$ plane, only the $s_x$ polarization associated with the $A_x$ mode is permitted; on the $k_y = 0$ plane, only the $s_y$ polarization originating from the primary $G_y$ order survives; and on the $k_z = 0$ plane, only the $s_z$ polarization associated with the $C_z$ order is allowed. 

Because FM canting is symmetry-forbidden, all allowed spin projections represent purely momentum-compensated spin textures without any isotropic background. For LaTiO$_3$ in the $G_y$ configuration (Fig.~\ref{fig:gamma4}(a)), the primary $G_y$ order enforces a pronounced, momentum-compensated $s_y$ spin polarization on the $k_y = 0$ plane, whereas the $s_x$ and $s_z$ projections on their respective symmetry-allowed planes remain extremely faint due to negligible $A_x$ and $C_z$ canting modes. Similarly, for $\text{CaCrO}_3$ in the $C_z$ configuration (Fig.~\ref{fig:gamma4}(b)), the dominant $C_z$ order dictates a strong, compensated $s_z$ spin polarization on the $k_z = 0$ plane, while $s_x$ and $s_y$ remain virtually zero owing to negligible $A_x$ and $G_y$ cantings. This clear correspondence between the first-principles calculations and Eq.~\eqref{eq:4gamma} validates the symmetry-enforced momentum constraints of the $\Gamma_4$ phase.

\subsection{\textbf{\boldmath{Anomalous Hall conductivity}}}
\label{subsec:ahcccc}

To validate our symmetry analysis in the context of the anomalous Hall conductivity (AHC), we investigate two representative compounds from Table~\ref{tab:allmaterials}: the Mott insulator LaTiO$_3$ and the metallic oxide CaCrO$_3$. Although their experimental ground states exhibit distinct magnetic orders---$G_z$-type in LaTiO$_3$~\cite{Meijer1999_LaTiO3} and $C_y$-type in CaCrO$_3$~\cite{Ofer2010_CaCrO3_muSR,Komarek2008_CaCrO3}---both belong to the same irreducible representation, $\Gamma_2$. Consequently, they share an identical magnetic symmetry, which permits only the $\sigma_{yz}$ component of the AHC tensor. While the AHC of CaCrO$_3$ has previously been reported in Ref.~\cite{Nguyen2023_CaCrO3_AHE}, our calculations employ a different choice of Wannier projections and significantly denser $k$-point sampling, resulting in few quantitative differences. Furthermore, in that work, time-reversal symmetry was not incorporated in the momentum-space symmetry analysis (see their Table~2)\cite{Nguyen2023_CaCrO3_AHE}, which likely led to incorrect little-group assignments reported in their Table~4.

Consistent with the symmetry analysis, we obtain sizable anomalous Hall conductivities exclusively in the symmetry-allowed $\sigma_{yz}$ component. As shown in Fig.~\ref{fig:ahc}(a,d), LaTiO$_3$ exhibits an AHC of approximately 38~S/cm at about 0.02~eV below the Fermi level, whereas CaCrO$_3$ reaches nearly 205~S/cm at the Fermi level. Remarkably, these substantial Hall responses arise despite their nearly vanishing net magnetizations, which are only 0.03~$\mu_B$/f.u. for LaTiO$_3$ and 0.25~$\mu_B$/f.u. for CaCrO$_3$, both oriented along the $x$ direction.

To elucidate the microscopic origin of the AHC, we examine the electronic band structures along momentum-space paths exhibiting large Berry curvature. In LaTiO$_3$, the avoided crossings immediately below the Fermi level are predominantly generated by spin--orbit coupling (SOC), as shown in Fig.~\ref{fig:ahc}(c). However, at energies around $\sim -0.2$~eV, the bands are already spin-split in the absence of SOC owing to the underlying altermagnetic symmetry (Fig.~\ref{fig:ahc}(b)). The inclusion of SOC transforms these pre-existing band crossings into small anticrossing gaps through band hybridization, thereby producing pronounced Berry-curvature hotspots. An analogous mechanism operates in CaCrO$_3$, where spin-polarized bands intersect near the Fermi level without SOC (Fig.~\ref{fig:ahc}(e)). Upon introducing SOC, these crossings are lifted to form anticrossing gaps (Fig.~\ref{fig:ahc}(f)), leading to a strong enhancement of the Berry curvature and, consequently, a large anomalous Hall conductivity.

These results demonstrate that the anomalous Hall response is not driven solely by the weak SOC-induced net magnetization but also arises from the non-relativistic spin splitting inherent to the altermagnetic electronic structure. The role of SOC is therefore primarily to gap the symmetry-protected band crossings, thereby generating large Berry curvature. In this regard, LaTiO$_3$ and CaCrO$_3$ provide compelling evidence that altermagnetic materials can host substantial anomalous Hall effects even in the presence of negligible net magnetization. \\

\section{Summary and Conclusions}
In this work, we established a comprehensive symmetry framework for understanding altermagnetic spin splitting in orthorhombic $Pbnm$ perovskite antiferromagnets. Using spin group theory, we derived the symmetry-allowed momentum-dependent spin-splitting terms for the three fundamental collinear antiferromagnetic orders and identified the corresponding momentum-space planes where spin splitting is symmetry protected. Extending the analysis to the spin--orbit-coupled regime using magnetic group theory, we showed that the magnetic irreducible representation uniquely determines the effective low-energy Hamiltonian, the symmetry-allowed weak ferromagnetic canting, and the nonvanishing anomalous Hall conductivity tensor components.

These symmetry predictions were systematically validated by first-principles calculations on eight experimentally realized orthorhombic perovskite oxides, which accurately reproduced the characteristic plane-selective spin splitting and the symmetry-imposed electronic structure. Furthermore, calculations of the anomalous Hall conductivity for LaTiO$_3$ and CaCrO$_3$ revealed sizable Hall responses despite nearly vanishing net magnetization. Our analysis demonstrates that the anomalous Hall effect originates primarily from the non-relativistic altermagnetic spin splitting, while spin--orbit coupling plays the essential role of opening gaps at symmetry-protected band crossings, thereby generating large Berry curvature.

Beyond identifying individual material candidates, this work establishes magnetic symmetry as the fundamental organizing principle connecting magnetic order, electronic structure, and Hall transport in orthorhombic altermagnets. The symmetry-guided framework developed here provides a predictive strategy for discovering new altermagnetic materials and engineering spin-dependent electronic and transport functionalities in antiferromagnetic spintronic devices.

\bibliographystyle{unsrt}
\bibliography{draft1}
\end{document}